\documentclass[pteplogo]{ptephy_v1}
\usepackage{graphicx}
 \newcommand{\VEC}[1]{\vec {#1} }
 \def\Alfven{Alfv{\'{e}}n\ }
\begin{document}
\title{The growth of chiral magnetic instability
in a large-scale magnetic field}
 \author{\name{Yasufumi Kojima}{} and \name{Yuri Miura}{}}
\address{
\affil{}{Department of Physics, Hiroshima University, 
Higashi-Hiroshima, Hiroshima 739-8526, Japan}
\email{ykojima-phys@hiroshima-u.ac.jp}}
\begin{abstract}
The chiral magnetic effect emerges from a miroscopic level, and 
its interesting consequences have been discussed in
the dynamics of the early universe, neutron stars and quark-gluon plasma. 
An instability is caused by anomalous electric current 
along magnetic field.
We investigate effects of plasma motion on the instability
in terms of linearized perturbation theory.
A magnetic field can inhibit magnetohydrodynamic waves
to a remarkable degree and thereby affects the instability mode.
We also found that the unstable mode
is consisted of coupling between \Alfven  and one of magneto-acoustic waves.
Therefore, the propagation of a mixed \Alfven  wave 
driven by magnetic tension is very important.
The direction of unperturbed magnetic field favors
the wave propagation of the instability mode,
when \Alfven speed exceeds sound speed.
\end{abstract}
%
\subjectindex{ MHD : dynamo : neutron stars : early universe}

\maketitle
\section{Introduction}

It is widely recognized that the magnetohydrodymamics (MHD)
is capable of describing a variety of astrophysical phenomena.
The treatment is macroscopic one consisted of
>fluid motions coupled with electromagnetic forces.
A set of MHD equations is scale-independent, and
may be applied to the laboratory to astrophysical plasma.
The electromagnetic fields are described by classical Maxwell's 
equations. They have a symmetry with respect to a parity,
that is, a transformation property under spatial inversion.
Physical vectors can equate only vectors of the same kind.
An example is Ohm's law ${\VEC j} =\sigma {\VEC E}$ with a scalar $\sigma$, 
where electric field ${\VEC E}$ and electric current ${\VEC j}$ are 
polar vectors. Magnetic vector ${\VEC B}$ is an axial one, and
is connected to the polar current vector ${\VEC j} $ as
${\VEC \nabla} \times  {\VEC B} = 4\pi {\VEC j}/c$
(e.g., \cite{1975clel.book.....J}).
A peculiar form of an electric current arises from a microscopic level:
\begin{equation}
{\VEC j} = \kappa {\VEC B} .
    \label{curntb.eqn}
\end{equation}
This form means that 
$\kappa$ is not scalar but pseudo-scalar by the parity transformation.
Possible origin of the form (\ref{curntb.eqn})
is a quantum anomaly known as the chiral magnetic effect
(e.g.,\cite{1980PhRvD..22.3080V,
1985PhRvL..54..970R,1998PhRvL..81.3503A,
2010PhRvL.104u2001F,2013PhRvL.111e2002A} and and references therein).
There is imbalance between left-handed and right-handed particles
in a quantum system, and the current flow along the magnetic field 
emerges in a macroscopic level.
It is known that the electric current along the magnetic field
causes an instability, leading to a growth of magnetic field
(e.g., \cite{1997PhRvL..79.1193J,2015PhRvD..92d3004B,
2016PhRvD..94b5009B}).
Magnetic helicity, which is an indicator of a global topology,
is also changed as well as the field amplification.
It has been discussed that the chiral magnetic effect
leads to inverse cascade, that is,
energy transfer from small to large scales.
The problem is studied from various aspects\cite{1999PhRvD..59f3008S,
2007PhRvL..98y1302C,2015PhRvL.114g5001B,2015PhRvD..92l5031H,
2016PhRvD..93l5016Y,2017PhRvD..96b3504P}.
This property is an important process of self-organization 
of turbulent structure.
In recent years, the chiral magnetic effect is widely discussed 
in relation to quark-gluon plasma in heavy-ion collision experiment
\cite{2016PrPNP..88....1K},
and astrophysical consequences in the early universe
\cite{1997PhRvL..79.1193J,1999PhRvD..59f3008S,2017ApJ...845L..21B},
core-collapse super-nova\cite{2016PhRvD..93f5017Y, 2018PhRvD..98h3018M},
magnetar\cite{2018MNRAS.479..657D}.
The electric current (\ref{curntb.eqn})
is also discussed in a context of mean-field MHD 
dynamo(e.g.,\cite{1978mfge.book.....M, 1980mfmd.book.....K,
1983flma....3.....Z}).
In the theory, the mean values of the variables
can be distinguished from the fluctuating ones.
Thus, the electric current (\ref{curntb.eqn}) on a large scale
is caused as an ensemble of screw-like vortices
in microscopic turbulence. 
Direct numerical simulations of MHD dynamo
have been developed with the increase of computer power.
In the approach, dynamics of all scales 
is simultaneously followed as far as small-scale waves
are numerically resolved. 
Thus, a model of microscopic turbulence 
is no longer needed there.
Indeed, some chiral MHD simulations have been performed 
in non-relativistic framework\cite{2017ApJ...846..153R,2018ApJ...858..124S}
and in relativistic framework\cite{2018MNRAS.479..657D}.
Their remarkable results demonstrate the ability of the method.
However computational cost may be high in the 
high resolution simulations.
Another example of the form (\ref{curntb.eqn})
is force-free magnetic fields
(e.g., \cite{1978mfge.book.....M,1996ffmg.book.....M}), in which
Lorentz force vanishes ${\VEC j} \times{\VEC B} =0$.
In a stationary case,  $ \kappa $ is constant along 
magnetic field line. The magnetosphere around
a star is modeled by the force-free approximation, in  
which magnetic pressure is assumed to be much larger than 
thermal one.
The macroscopic dynamics are governed by the same equations, 
although the transport coefficient $ \kappa $
determined by a microscopic process is various in the magnitude.
Bearing various astrophysical environments in mind,
it is important to explore the instability in a 
wide range of parameters.
In this paper, we consider normal mode analysis
for linearized system of chiral MHD equations.
The background state is assumed to be homogeneous with uniform 
magnetic field, and small perturbations
propagate as MHD waves in the absence of
chiral magnetic effect.
We study the modification of wave propagation and
the instability caused by the chiral magnetic effect in Section 2.
This problem was partially studied\cite{2017ApJ...846..153R},
where the modification is found, but the general property is not clear.
The reason will be discussed after our results.
We extensively analyze it to explore relevance of the instability in
various astrophysical environments.
We discuss our results in Section 3.

\section{Waves in a linearized system}
\subsection{Equations}


A set of chiral MHD equations are discussed in literature(e.g., 
\cite{2017ApJ...846..153R,2018ApJ...858..124S}).
The linear perturbation equations 
in non-relativistic dynamics are summarized here.
We assume that the unperturbed state of the medium is
static and homogeneous, i.e., the density $\rho_{0}$ 
and magnetic field ${\VEC B}= B_{0}{\VEC e}_{z}$, where
$\rho_{0}$ and $B_{0}$ are constant.
We write small perturbation 
of a quantity $Q$ as $\delta Q$, and then
the perturbation equations are given by
\begin{equation}
\frac{ \partial }{\partial t} \delta {\VEC B}
= - {\VEC \nabla} \times(c \delta {\VEC E} ),
    \label{Frad.eqn}
\end{equation}
\begin{equation}
c \delta {\VEC E} = -\delta {\VEC v} \times {\VEC B}_{0}
+\eta  {\VEC \nabla} \times \delta {\VEC B} - \kappa \delta {\VEC B} ,
    \label{Edef.eqn}
\end{equation}
\begin{equation}
\frac{ \partial }{\partial t} \delta \rho
+ {\VEC \nabla}  \cdot ( \rho_{0} \delta {\VEC v} )=0,
    \label{Evdns.eqn}
\end{equation}
\begin{equation}
\rho_{0} \frac{ \partial }{\partial t} \delta {\VEC v} 
= -  {\VEC \nabla}( c_{s}^2 \delta \rho) 
+ \frac{1}{4\pi} ( {\VEC \nabla} \times \delta {\VEC B})
 \times {\VEC B}_{0},
    \label{Evmot.eqn}
\end{equation}
where an adiabatic relation $ \delta p=  c_{s}^2 \delta \rho$ and
the {Amp{\`{e}}re's} law
$ 4\pi  \delta {\VEC j} =c {\VEC \nabla} \times \delta {\VEC B}$
are used. 
An electric current parallel to magnetic field is added 
by the chiral magnetic effect in eq.(\ref{Edef.eqn}).
We denote a sound speed as $ c_{s}$ and also 
\Alfven speed as 
$ c_{a} = (B_{0}^2/( 4 \pi \rho_{0}) )^{1/2}$.
There are two kinds of restoring forces,
pressure and magnetic tension on the plasma motion.
Relative importance is inferred from the ratio,
which corresponds to so-called plasma $\beta$;
$\beta \approx (c_{s}/c_{a})^{2}$.  
We assume that electric resistivity $\eta$ and 
a coefficient $\kappa$ of chiral magnetic effect are constant.
The latter is determined by chiral chemical potential, i.e.,
imbalance between left and right-chiral particles.
As the chiral magnetic instability grows, 
an electric current flows along magnetic field, and 
the magnitude $\kappa$ eventually reduces.
Our concern is the linear growth at the initial
stage, so that $\kappa$ is regraded as a constant.
We assume that all perturbed quantities are proportional to
the Fourier form $\exp(-i(\omega t -{\vec k} \cdot {\vec x}))$, 
and that the wave propagates in $x$-$z$ plane, i.e.,
$ {\VEC k} = ( k_{x}, 0, k_{z})$
$ = ( k \sin \theta, 0, k \cos \theta )$.
We also assume $k > 0$, but
the frequency $\omega$ in general is a complex number,
$\omega \equiv \omega_{\rm R} +i \omega_{\rm I}$. 
A mode  with $\omega_{\rm I} >0$ grows with time.
We may limit to the case of $\kappa \ge 0 $, since the
chiral instability depends on $ \kappa^2$ as shown below, although
both signs of $\kappa$ are physically allowed.
After some manipulation, perturbed equations 
(\ref{Frad.eqn})-(\ref{Evmot.eqn}) are reduced to 
\begin{equation}
 M \delta {\bf u} = 0 ,
\end{equation}
where $M$ is a $5 \times 5$ matrix whose 
components are explicitly given by
\begin{equation}
M=\left[ \begin{array}{ccccc}
k_{z} & 0 & 0   &
 \omega +i \eta k^2 & - k_{z} \kappa \\
0 & k_{z} & 0   &
 \kappa k^2 k_{z} ^{-1}  & \omega +i \eta k^2 \\
\omega^2 - c_{s}^2 k_{x}^2 & 0 & - c_{s}^2 k_{x} k_{z} & 
\omega c_{a}^2 k^2 k_{z}^{-1} & 0 \\
0 & \omega & 0 &
0 &  c_{a}^2 k_{z}\\
- c_{s}^2 k_{x} k_{z} & 0 & \omega^2 - c_{s}^2 k_{z}^2 &
0 & 0 \\
\end{array} \right],
\end{equation}
and a vector $\delta {\bf u} $ is given by
$\delta {\bf u} ^{\rm T}=(\delta v_{x}, \delta v_{y}, \delta v_{z},
B_{0}^{-1} \delta B_{x}, B_{0}^{-1} \delta B_{y})$.
The component $\delta B_{z}$ is determined by the Gauss law
${\VEC \nabla} \cdot \delta {\VEC B}=0$, i.e,
$\delta B_{z}= -\delta B_{x} k_{x}  k_{z}^{-1} .$
A determinant of the matrix $M$ provides a dispersion relation:
\begin{eqnarray}
\nonumber
0 & =& 
(\omega ^2 -  c_{s}^{2} k^2) \omega ^2 X^2
-[ \omega ^2 ( k_{x} ^2 + 2k_{z}^2) -2 c_{s}^{2}k_{z}^2 k^2]c_{a}^{2} \omega X \\
& &
+ [ \kappa^2 \omega^4
  +(c_{a}^{4} k_{z} ^2 -\kappa^2 c_{s}^{2} k^2) \omega^2
- c_{a}^{4}  c_{s}^{2} k_{z}^4 ]k^2 ,
    \label{dispersion.eqn}
\end{eqnarray}
where $ X= \omega +i\eta k^2$. Equation (\ref{dispersion.eqn}) is 
a sixth degree polynomial in $\omega$.
In order to understand the general feature of the solutions,
we start with some limiting cases in next subsections.
%

\subsection{Chiral magnetic instability}
We first consider the case of $c_{s}=c_{a}=0$.
That is, there are no forces on plasma motions.
The dispersion relation (\ref{dispersion.eqn}) becomes
\begin{equation}
0 = [(\omega +  i \eta k^2) ^2 + \kappa^2 k^2 ]\omega^4.
\end{equation}
Non-zero solution is given by $\omega = - i \eta k^2 \pm i \kappa k$.
The mode with $ \omega = -i(\kappa k +\eta k^2)$ always decays.
However, the mode with $ \omega = i( \kappa k-  \eta k^2)$
grows for $ k < \eta ^{-1} \kappa$.
That is, the long wavelength mode with
 $ \lambda  > \lambda_{c} \equiv |\kappa|^{-1} \eta$
is unstable. Resistivity is inefficient for such a long wavelength mode.
Eigenvectors of perturbation functions satisfy
\begin{equation}
\delta {\VEC v}= (0,~0,~0),
~~
\delta  {\VEC B} \propto (k_{z},  ~\pm i k , ~k_{x}).
\end{equation}
These functions mean that the disturbance is purely magnetic one, and
is transverse to the wave vector, i.e.,
${\VEC k} \cdot \delta {\VEC B} =0 $ and also
${\VEC k} \cdot \delta {\VEC j} =0 $.
Our concern is the chiral instability mode due to $\kappa \ne 0$, 
so that we from now on neglect the resistivity $\eta =0$.
The approximation is valid in the long wavelength mode
$\lambda \gg \lambda_{c}$.
This approximation simplifies the
dispersion relation (\ref{dispersion.eqn}), 
which is reduced to a cubic equation of $\omega^2$.
%

\subsection{Effect of thermal pressure}
By setting  $\eta =0$ and $ c_{a}=0$ in eq.(\ref{dispersion.eqn}),
we have a relation:
\begin{equation}
0 = (\omega^2 + \kappa ^2  k^2)(\omega^2 -c_{s}^2 k^2)\omega^2 .
\end{equation}
There are two non-trivial solutions. 
One is chiral magnetic mode ($\omega^2 =-\kappa ^2 k^2$), 
and the other sound wave mode ($\omega^2 =c_{s}^2 k^2 $).
The sound wave is produced by compressional motion of matter
and hence longitudinal mode, i.e.,
 $ {\VEC k} \cdot \delta  {\VEC v} \ne 0$. 
 We explicitly check this fact by the eigen-functions: 
\begin{equation}
 \delta {\VEC v}   \propto  {\VEC k} =( k_{x},~ 0,  ~k_{z} ),
~~ \delta {\VEC B } = 0.
\end{equation}
As discussed in previous subsection, 
the chiral mode is transverse mode
$ {\VEC k} \cdot \delta  {\VEC B} =0$.
These two modes are completely decoupled.
The growth rate of chiral magnetic mode
is not affected by the pressure.
%

\subsection{Effect of a uniform magnetic field}
In the case of  $c_{s}=0$ and $\eta =0$, 
the dispersion relation (\ref{dispersion.eqn}) is reduced to
\begin{equation}
0 = [(\omega^2 -c_{a}^2 k_{z}^2)(\omega^2 -c_{a}^2 k^2)
 + \kappa ^2  k^2 \omega ^2] \omega ^2 .
    \label{dscagn.eqn}
\end{equation}
It is clear that two waves, the \Alfven mode and fast MHD mode
(or fast magneto-acoustic mode) are coupled by a $\kappa$-term.
Frequency of the slow one is zero ($ \omega ^2 = 0$) 
in the limit of $c_{s}=0$.
Two non-zero solutions are given by
\begin{equation}
\frac{\omega ^2}{k^2} 
=\frac{1}{2}\left[(1+\cos^2 \theta )c_{a}^2 -\kappa^2 
\pm Q^{1/2} \right],
    \label{AlfMagwv.eqn}
\end{equation}
where
\begin{equation}
Q=\left[ (1-\cos\theta)^2c_{a}^2-\kappa^2 \right]
  \left[(1+\cos\theta)^2c_{a}^2-\kappa^2  \right] .
\end{equation}
The function $Q$ becomes negative in a range of 
\begin{equation}
\frac{1}{(1+|\cos\theta|)^2} < \frac{c_{a}^2}{\kappa^2}
< \frac{1}{(1-|\cos\theta|)^2} ,
    \label{CondCaQ.eqn}
\end{equation}
and hence $\omega $ becomes a complex number.
Outside the range of eq.(\ref{CondCaQ.eqn}), 
$\omega $ is either real or pure imaginary.
Nature of a solution $\omega$ is thus classified to three regions
in the $c_{a} \kappa^{-1}$ - $\theta $ plane, as shown in Fig.1.
In the region I (left part of the figure),
where $ c_{a} \kappa^{-1} <( 1+|\cos\theta|)^{-1}$,
the solution is a pure imaginary i.e., $\omega_{\rm R}=0 $.
On the other hand, $\omega $ is real ($\omega_{\rm I}=0 $),
in the region III (the up-right part of the figure).
That is stable wave region,
in which $ c_{a} \kappa^{-1} >(1-|\cos\theta|)^{-1}$.
In the intermediate region II, the frequency is a complex number, 
$\omega = \omega_{\rm R}+ i \omega_{\rm I} $.
The mode becomes oscillatory instability.
%

\begin{figure}[bt]
\begin{center}
  \includegraphics[scale=1.0]{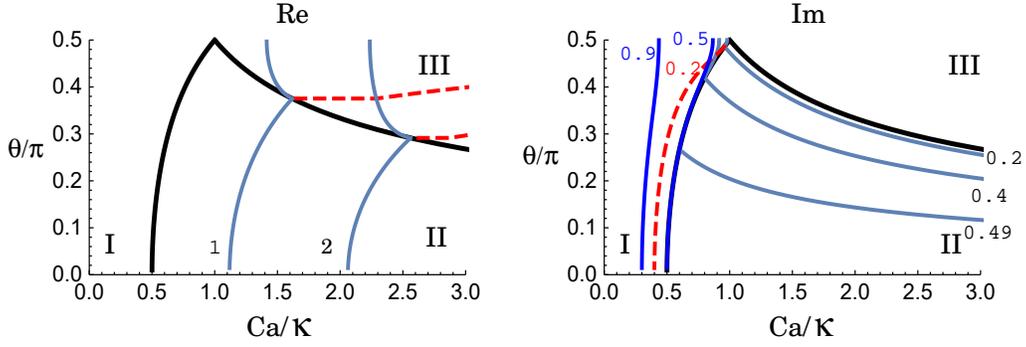}%
\caption{ Contours of $\omega_{\rm R}/(k \kappa)$(left panel) 
and $\omega_{\rm I}/(k \kappa)$(right panel)
in the  $c_{a} \kappa^{-1}$ - $\theta $ plane.
In the region I of the left panel, $\omega_{\rm R}/(k \kappa)$ is 0, but
$\omega_{\rm R}/(k \kappa)$ increases 
with $c_{a}\kappa^{-1}$ in the region II.
Constant lines with $\omega_{\rm R}/(k \kappa) =1,2$ are plotted.
In region III, there are two stable waves, fast MHD
and \Alfven waves for a fixed velocity $\omega_{\rm R}/(k \kappa)$.
In the region I of the right panel, there are two kinds of growing modes. 
In region III of the right panel, $\omega_{\rm I}/(k \kappa) $ is zero.
Some constant lines are plotted
for values $\omega_{\rm I}/(k \kappa) $ labeled in the figure.
}
\end{center}
\end{figure}
%

\begin{figure}[bht]
\begin{center}
  \includegraphics[scale=0.8]{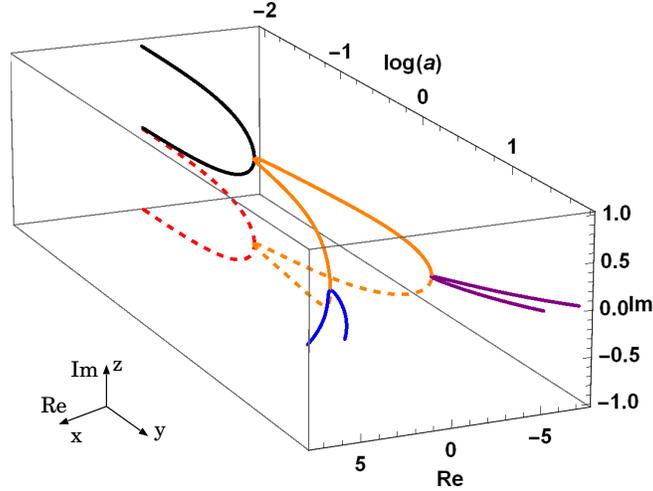} %
\caption{ Three dimensional display of mode
coupling as a function of  $\log_{10}(c_{a}\kappa^{-1})$,
which is chosen as $y$ axis ($-2 \le \log_{10}(c_{a}\kappa^{-1}) \le 2$).
Real and imaginary parts of a mode are shown in
$x$ ($-8 \le {\rm Re}( \omega/k \kappa) \le 8$)
and $z$ ($-1 \le {\rm Im}( \omega/k \kappa) \le 1$) axes.
Negative value of the imaginary part is 
plotted by a dashed line.  
For large $c_{a}\kappa^{-1}$, two
modes are stable waves,  fast MHD and \Alfven waves.
}
\end{center}
\end{figure}
%

In Fig.1, we demonstrate some contours of real and imaginary parts
as a function of $ c_{a} \kappa ^{-1}$ and propagation angle $\theta$.
We may limit to the case of $\omega_{\rm R} >0 $ and $\omega_{\rm I} >0 $ only, 
since a pair of  $ \pm (\omega_{\rm R} + i\omega_{\rm I} ) $
is always a solution.
The real part $\omega_{\rm R}/(k \kappa)$, 
a normalized phase velocity is zero in the region I 
of the left panel, but it
increases with $c_{a}\kappa^{-1}$ in the region II
for a fixed angle $\theta$.
There are two solutions in region III.
They are identified as fast MHD and \Alfven waves. 
The former is approximated as
$ (\omega/k)^{2} \approx c_{a}^2 - (\kappa / \sin \theta)^{2}$
for $c_{a}\kappa^{-1} \gg 1 $, and
$\omega_{\rm R}/(k \kappa)$ does not so strongly depend on $\theta$.
Therefore, a curve with constant velocity $\omega_{\rm R}/(k \kappa)$ 
becomes almost vertical in the left panel of Fig.1.
The other curve plotted by a horizontal dotted line represents
the  \Alfven wave, which is expressed as
$ (\omega/k)^{2} \approx c_{a}^2 \cos^2 \theta - \kappa ^{2} \cot^2 \theta $
for  $c_{a}\kappa^{-1} \gg 1 $.
%

Next we discuss the imaginary part $\omega_{\rm I}/(k \kappa)$.
For small $ c_{a} \kappa^{-1} $, two solutions in 
eq.(\ref{AlfMagwv.eqn}) are approximated as
\begin{equation}
\frac{\omega ^2}{k^2} =
\left\{ \begin{array}{l}
- \kappa ^{-2} c_{a}^4 \cos^4 \theta  + \cdots \\
 -\kappa ^2 + (1+\cos^2\theta ) c_{a}^2 + \cdots 
\end{array}
\right. 
    \label{cakplimt.eqn}
\end{equation}
where two functional forms correspond to the upper and lower
signs in eq.(\ref{AlfMagwv.eqn}).
There are two growing modes, and their characteristic growth rates 
are $\omega_{\rm I} \approx  c_{a}^2 \kappa^{-1} k$
(slowly growing mode), and  $\omega_{\rm I} \approx \kappa k$
(rapidly growing mode).
The frequency of the former vanishes in the limit of $ c_{a}=0$.
As $c_{a} \kappa^{-1}$ increases, both growth rates approach each other, 
and match on the critical line
$ c_{a} \kappa^{-1} =( 1+|\cos\theta|)^{-1}$.
In the right panel of Fig.1, some lines with 
constant $\omega_{\rm I}/(k \kappa)$ are plotted.
In the region I, there are two branches,
which are approximated by eq.(\ref{cakplimt.eqn}). 
In intermediate region II, the growth rate 
strongly depends on the propagation angle $\theta$. 
For example, the wave perpendicular to unperturbed 
magnetic field, i.e., $\theta =\pi/2$ 
is stabilized for $ c_{a} \kappa^{-1} \ge 1$.
On the other hand, the wave parallel to the magnetic field 
is never stabilized for any value of $ c_{a} \kappa^{-1} $.
%

We show the coupling of the \Alfven and fast MHD modes, 
which causes an unstable mode for $c_{a} \kappa^{-1} <1$.
Figure 2 displays how the phase velocity $ \omega/k$
normalized by $\kappa$ changes with the \Alfven velocity 
$c_{a} $, for fixed propagation angle $\theta =\pi/4$.
In the large limit of $c_{a} \kappa^{-1} $,
there are two different modes, which are described by 
positive velocities.
There are also negative velocity modes, but they are
physically the same as positive ones.
These different modes represent stable \Alfven and fast MHD waves. 
As $c_{a} \kappa^{-1} $ decreases, two velocities
agree at a certain point($c_{a} \kappa^{-1}\approx 3 $),
where two modes convert to one oscillatory growing 
and one oscillatory decaying modes in a region of $c_{a}\kappa^{-1} < 3$. 
The velocity $ \omega_{\rm R}/(k \kappa)$ further goes to 0, and
$\omega_{\rm R} =0$ at $c_{a}\kappa^{-1} \approx 0.6 $. 
Two propagating waves merge and change as standing waves
for $c_{a}\kappa^{-1} < 0.6$. 
Toward $c_{a}\kappa^{-1}  \to 0$ after that point, 
the frequencies change as 
$\omega_{\rm I} \to \pm 0 $ or $\omega_{\rm I} \to \pm  k\kappa $
with $\omega_{\rm R} = 0 $. 
The perturbation amplitudes satisfy
\begin{eqnarray}
&&
\delta v_{x}= - \frac{ c_{a}^2 k}{ \omega \cos\theta }
\frac{ \delta B_{x} }{ B_{0} },
~~
\delta v_{y}= - \frac{ c_{a}^2 k \cos\theta }{ \omega} 
\frac{ \delta B_{y} }{ B_{0} },
~~
\delta v_{z}=0,
\nonumber
\\
&&
(\omega ^2 - c_{a}^2 k^2) \delta B_{x} 
= \kappa \omega  k\cos\theta \delta B_{y},
~~
 \delta B_{z} = - \tan \theta \delta B_{x} .
    \label{eignvctA.eqn}
\end{eqnarray}
The perturbation of the magnetic field
is always perpendicular to the wave, since
$ {\VEC k } \cdot \delta {\VEC B} $
$ \propto {\VEC \nabla } \cdot \delta {\VEC B} =0$.
In order to study the direction of the plasma motion,
we consider two limiting cases.
For the wave parallel to the magnetic field ($\theta =0$),
plasma motion is also transverse to wave propagation,
since $\delta v_{z} =0 $.
There is no compression of matter,
$  {\VEC \nabla } \cdot \delta {\VEC v} =0$ in this case.
As $\theta \to \pi /2$,
we have $\delta v_{y} \to 0 $  as well as $\delta v_{z} =0 $.
This means that plasma motion is longitudinal,
$\delta {\VEC v} \propto {\VEC k } $, and is compressional 
$  {\VEC \nabla } \cdot \delta {\VEC v} \ne 0$ in this case. 
At intermediate angles of wave propagation,
the instability mode is a mixture of 
properties of two limiting cases.
%

\subsection{Magnetohydrodyamical effects}
In previous subsections, we have separately 
considered the effects of plasma motion driven by thermal 
pressure or magnetic tension on the chiral instability.
We here consider a combined effect by $c_{a} \ne 0$ and $c_{s} \ne 0$.
The dispersion relation (\ref{dispersion.eqn}) is
a cubic equation of $\omega^2$,
so that a pair ($\pm \omega$) is always a solution of it.
It is also easy to understand the fact that there is at least  
one solution of $\omega^2 >0 $, i.e, a stable wave.
  Figure 3 shows the maximum growth rate  $\omega_{\rm I}/(k \kappa)$ 
among four solutions in $c_{s}\kappa^{-1} $ - $c_{a}\kappa^{-1} $ plane,
for the propagation angle $\theta =\pi/4$.
Stable wave-propagation region is expressed by upper part 
of `$\nu$'-shape.
It is natural that there is a different nature
that depends on the dominant force.
In high $\beta$ region(lower right part of Fig.3), 
there is an unstable mode.
The growth rate is $\omega_{\rm I}/(k \kappa) \approx 1 $ in
the limit of $c_{a}\kappa^{-1} =0$, irrespective of $c_{s}\kappa^{-1}$.
Plasma motion driven by dominant pressure force
does not affect the instability, as discussed in subsection 2.3.
As $c_{a}\kappa^{-1}$ increases
with a fixed value of $c_{s}\kappa^{-1} (> 1.5)$, 
$\omega_{\rm I}/(k \kappa) $ decreases and becomes 0 at
$c_{a}\kappa^{-1} \approx c_{s}\kappa^{-1}$.
However, unstable bound of $\omega_{\rm I}>0$ shifts 
to a larger value of $c_{a} \kappa^{-1}$ for a small
$c_{s} \kappa^{-1} $, that is, a fat part of Fig.3,
where $c_{a}\kappa^{-1} > c_{s}\kappa^{-1}$.
%

\begin{figure}[bht]
\begin{center}
  \includegraphics[scale=1.20]{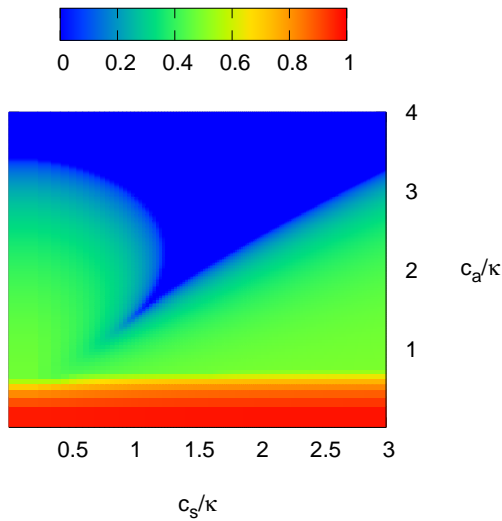} %
\caption{ Color contour of maximum growth rate ${\rm Im} (\omega/(k \kappa))$
in $c_{s}\kappa^{-1}$-$c_{a}\kappa^{-1}$ plane.
The propagation angle of the perturbation is $\theta =\pi/4$. 
All modes become stable propagation waves in upper `$\nu$'-shaped region
(a blue region in the figure).
}
\end{center}
\end{figure}
%

In order to study the unstable mode, we show  in Fig.4 
how the frequency of a mode changes with $c_{a}\kappa^{-1}$ 
for a fixed $c_{s}\kappa^{-1}$.
In the case of $c_{s}\kappa^{-1} =0.75$ (left panel of Fig.4),
all modes become stable waves for $c_{a}\kappa^{-1} \ge 3$.
Their phase velocities $ \omega/k$ characterize
the waves, so that we identify the fast MHD, \Alfven and
slow MHD waves according to the absolute value of velocity.
It is also found that the unstable mode
is caused by a coupling of the fast and \Alfven waves
as $c_{a}\kappa^{-1} $ decreases.
The slow one is always decoupled, and is stable wave.
This situation is the same as that considered in previous
subsection ($c_{s}=0$), where the slow mode 
is decoupled as $\omega=0$. 
%

In the right panel of Fig.4,
we show the case of $c_{s}\kappa^{-1} =2$.
Like the previous case, three stable waves are 
identified for $c_{a}\kappa^{-1} \ge 3$.
As $c_{a}\kappa^{-1} $ decreases,
the coupling occurs between the slow and \Alfven modes.
The fast mode is always decoupled.
This point differs from that for $c_{s}\kappa^{-1} =0.75$.
Unstable region in Fig.3 changes
by a MHD mode which the \Alfven mode couples with.  
The \Alfven mode couples with slow one in a high $\beta$ region
($ c_{s} > c_{a}$), whereas it couples with fast one in a 
low $\beta$ region ($ c_{a} > c_{s}$).
It is interesting to observe the behavior in
a small $ c_{a}\kappa^{-1} $ region in the left panel of Fig.4
(the case of  $c_{s}\kappa^{-1} =0.75$).
The phase velocity $\omega_{\rm R}/(k \kappa)$ of the instability mode
sharply decreases around $ c_{a}\kappa^{-1} = 0.75$.
At that point, velocity of slow mode sharply increases.
That is, wave nature is exchanged.
The unstable mode originates from a coupling of fast and \Alfven waves
in a low $\beta$ region, but
the nature changes like slow one in a high $\beta$ region.
At the same time, stable mode behaves like the slow one in 
large $ c_{a} $ region, but behaves like the fast one
in small $ c_{a} $ region.
%

\begin{figure}[bht]
  \includegraphics[scale=0.8]{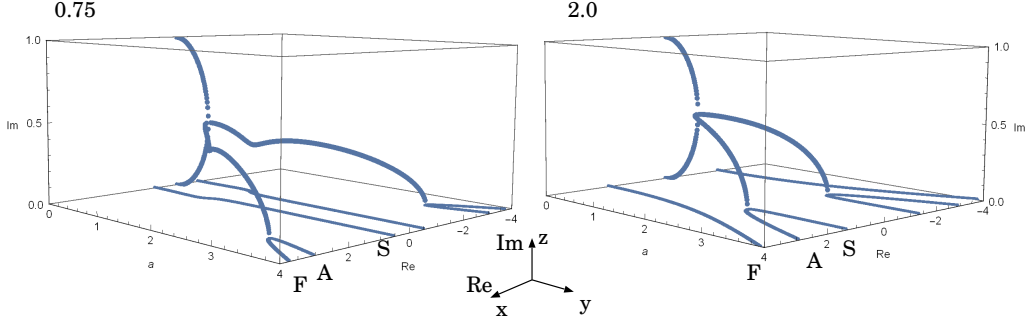} %
\caption{Three dimensional display of mode
coupling as a function of  $c_{a} \kappa^{-1}$,
which is chosen as $y$ axis ($0 \le c_{a} \kappa^{-1} \le 4$).
Real and imaginary parts of a mode are shown in
$x$ ($-4 \le {\rm Re}( \omega/k \kappa) \le 4$)
and $z$ ($0 \le {\rm Im}( \omega/k \kappa) \le 1$) axes.
Left panel is for $c_{s} \kappa^{-1} = 0.75$, while
right one is for $c_{s} \kappa^{-1} = 2$.
For large $c_{a} \kappa^{-1}$, all modes are stable
waves characterized by a real frequency.
By their propagation velocity, they are identified
as F(fast MHD), S(slow MHD) and 
A(\Alfven) waves, as labeled in the figure. 
}
\end{figure}
%

\begin{figure}[bth]
\begin{center}
  \includegraphics[scale=1.0]{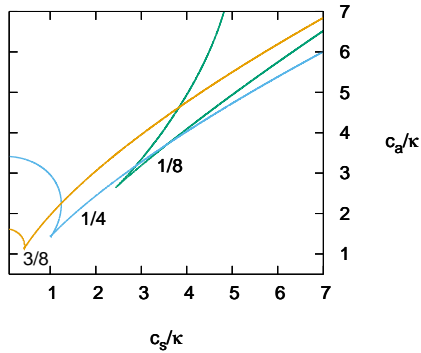} %
\caption{Stable wave region in a parameter space
of $c_{s}/\kappa$ and $c_{a}/\kappa$.
A region between two curves of `$\nu$'-shape 
denotes stable wave propagation for a fixed angle.
Three curves are plotted for $\theta =\pi/8, \pi/4, 3\pi/8$.
Stable wave region increases with the increase of propagation angle 
$\theta$.
}
\end{center}
\end{figure}
%

We discuss how the stable wave region changes with the 
propagation angle $\theta$. Figure 5 shows the region 
for $\theta =\pi/8, \pi/4$ and $3 \pi/8$.
The instability is almost unchanged in a high $\beta$ region
($c_{s} \kappa^{-1}> c_{a}\kappa^{-1}$), since $c_{a}$ is unimportant.
However, the growth rate significantly depends on the angle $\theta$
in a low $\beta$ region ($c_{a} \kappa^{-1}> c_{s}\kappa^{-1}$), 
where the \Alfven wave propagation affects the instability.
As discussed in subsection 2.4,
unstable region diminishes with the increase of $\theta$ 
for $c_{a} \kappa^{-1} >1 $. 
The \Alfven mode velocity goes to $0$ in 
orthogonal direction to unperturbed magnetic field, $\theta \to \pi/2$.
Accordingly, the growth is suppressed in a low $\beta$ region with 
$c_{a} \kappa^{-1} >1 $.
A peculiar thing should be noted
in the limit of $\theta =\pi/2$ ($\cos \theta =0$).
The behavior of $\cos \theta =0$ differs from that 
of $\cos \theta \approx 0$.
The dispersion relation for $\cos \theta =0$
is analytically expressed, and shows
that there is always one growing mode, for
any values of $c_{a} \ne 0 $ and $c_{s} \ne 0 $.
In an exactly perpendicular direction,
the \Alfven wave propagation is prohibited,
and the instability grows irrespective of 
magnetic field strength.
%

\section{Summary and Discussion}

%
The chiral magnetic instability is inherent in an electromagnetic 
field with the electric current parallel to the magnetic field.
We have taken into account of the plasma motion
in order to study its relevance in various environments.
Assuming that disturbances are small, linearized equations of chiral MHD 
are examined.  All modes are described by solving the resultant 
dispersion relation, no matter whether they are stable or unstable. 
The analysis of it is not new, as mentioned in Introduction.
We should therefore discuss previous results\cite{2017ApJ...846..153R}, 
and especially insufficient points of previous analysis.
The dispersion relation depends on three parameters:
a set of independent ones is chosen as 
$c_{a}/\kappa$, $c_{s}/\kappa$ and $\theta$ in this paper.
In Ref. \cite{2017ApJ...846..153R}, 
they fixed a ratio of $c_{a}/c_{s}$, and plotted the phase 
velocity of MHD waves as a function of propagation 
angle $\theta$ for $(c_{s}/\kappa )^2=0.1,1,10$.
We found that the choice is not good to grasp whole structure of 
normal frequencies, as inferred from Fig.3.
Their parameters are also limited to stable wave region, and 
growth rates are never discussed.
The relation to the instability is unclear, although modified MHD waves 
are shown by weak chiral magnetic effect.
Our choice is fixing the propagation angle at first, and all normal 
frequencies were calculated in a wider parameter space.
The growth rates were illustrated in three dimensional representation.
Thus, we have successfully explored the mode coupling in the unstable region.
We next summarize our findings.
When chiral magnetic effect is small enough, small disturbances are 
described by three waves, i.e, \Alfven, fast and slow MHD waves.
An unstable mode originates from a coupling of the \Alfven 
and one of magneto-acoustic waves.
The coupling condition is determined by matching the phase velocities, and
therefore the counterpart is slow one for 
high $\beta$ plasma ($c_{s} > c_{a} $), and fast one for
low $\beta$ plasma ($c_{s} < c_{a} $).
Astrophysically, the high $\beta$ plasma is relevant to
the early universe, and core-collapse super-nova and neutron star,
while low $\beta$ plasma is relevant to a force-free magnetosphere.
The \Alfven wave plays a very important role, since 
the magnetic perturbations are always transverse waves,
both in pure \Alfven mode and in pure chiral mode. 
%

The unstable mode grows regardless of the propagation direction
in high $\beta$ plasma, where pressure is a dominant force and
does not hinder the growth.
As a value of $\beta \approx (c_{s}/ c_{a})^2$ decreases, 
magnetic tension becomes important force on plasma motion.
Accordingly, wave propagation velocity depends on the direction.
In a low $\beta$ regime with $c_{a} > \kappa$,
three stable waves like in ordinary MHD appear
by mismatching their phase velocities.
The propagation of unstable mode perpendicular to the background
magnetic field is strongly constrained, i.e.,
disturbances propagate as stable waves in the direction.
In this way, we found that
the wave propagation and growth of the unstable mode
are significantly affected 
in the presence of the magnetic field on a large scale. 
Especially, in a low $\beta$-regime, the direction parallel to the field
is favored for unstable wave propagation, that is,
the instability anisotropically grows.
The situation may be related with structure formation with a larger 
coherent length of magnetic field, but
the issue is a non-linear process and is beyond the scope of this paper.
%


\end{document}